\documentclass[floatfix,aps,prl,twocolumn,showpacs,groupedaddress]{revtex4}

\usepackage{epsfig}

\newcommand{\beq}{\begin{equation}}
\newcommand{\bea}{\begin{eqnarray}}
\newcommand{\eeq}{\end{equation}}
\newcommand{\eea}{\end{eqnarray}}

\begin{document}

\title{Prethermalization}

\author{J.\ Berges}
\author{Sz.\ Bors\'anyi}
\author{C.\ Wetterich}
\affiliation{Institute for Theoretical Physics, Heidelberg University\\
Philosophenweg 16, 69120 Heidelberg, Germany\\}

\begin{abstract}
\noindent
Prethermalization of the equation of state and the
kinetic temperature to their equilibrium values
occurs on time scales dramatically shorter than
the thermal equilibration time. This is a crucial
ingredient for the understanding of collisions of
heavy nuclei or other nonequilibrium phenomena
in complex quantum and classical many
body systems. We also compare the chemical
equilibration time with other characteristic
time scales.
\end{abstract}
\pacs{11.10.Wx,12.38.Mh,05.70.Ln}

\maketitle

Prethermalization is a universal far-from-equilibrium phenomenon
which describes the very rapid establishment of an almost
constant ratio of pressure over energy density (equation
of state), as well as a kinetic temperature based on
average kinetic energy. The phenomenon occurs on time scales 
dramatically shorter than the thermal equilibration time.
As a consequence, prethermalized quantities approximately
take on their final thermal values already at a time
when the occupation numbers of individual momentum 
modes still show strong deviations from the late-time
Bose-Einstein or Fermi-Dirac distribution. 

The abundance of experimental data on matter in extreme conditions from
relativistic heavy-ion collision experiments, as well as applications in
astrophysics and cosmology urge a quantitative understanding 
of nonequilibrium dynamics in quantum field theories.
Collision experiments seem to indicate early thermalization
whereas the present theoretical understanding of QCD suggests
a much longer thermal equilibration time. For example, the
successful application of hydrodynamics already less than $1\,$fm
after the collision is so far 
unexplained from theory~\cite{Rapp:2004ki}. 

To resolve these questions, it is important to understand  
to what ``degree'' thermalization is required to explain the 
observations. Different quantities effectively thermalize 
on different time scales and a complete thermalization of all 
quantities may not be necessary. For instance,
an approximately time-independent 
equation of state $p=p(\epsilon)$, characterized by an almost 
fixed relation between pressure $p$ and energy density $\epsilon$, 
may form very early --- even though the system is still 
far from equilibrium! Such an almost constant equation of state
is a crucial ingredient for the use of efficient hydrodynamic 
descriptions, since it is needed to close the system of equations
obtained from the conservation of the energy momentum
tensor.  

The initial stages of a collision require to consider quantum fields
in extreme nonequilibrium situations.
Connecting this far-from-equilibrium dynamics at early
times with the approach to thermal equilibrium at late
times is a challenge for theory. Achieving this goal is crucial 
for a comparison between the time scales of prethermalization
and thermal equilibration, and we investigate this question here
quantitatively in an effective quark-meson model.
Approaches based on small deviations from equilibrium,
or on a sufficient homogeneity in time underlying kinetic descriptions,
are not applicable to describe the ``link'' between the initial
and the late-time behavior. Classical field theory
approximations are expected to be valid for not too late times, 
but cannot determine the relevant time scale for the approach 
to quantum thermal equilibrium. Recently, it has been 
demonstrated~\cite{Berges:2000ur,Cooper:2002qd,Aarts:2001yn,Berges:2002cz,Berges:2002wr} 
that far-from-equilibrium dynamics 
as well as subsequent thermalization of quantum fields can be described 
using efficient functional integral techniques. The 
description includes direct scattering as well as off-shell and memory 
effects. This is crucial to establish the different
time scales for a loss of memory of initial conditions for certain
``bulk quantities'' as compared to ``mode quantities'' characterizing 
the evolution of individual momentum modes.

The observation that the nonequilibrium system looses
a major part of the memory of the detailed initial conditions
on a very short time scale is a robust feature of classical
as well as quantum field theories. 
It has been observed~\cite{Bonini:1999dh} that approximate 
``equipartition'' between bulk kinetic and potential 
energy occurs very rapidly. Below, this will be the basis
of our definition of a ``kinetic temperature'' $T_{\rm kin}$.
We will see that the equation of state becomes almost
constant at the same ``prethermalization time'' $t_{\rm pt}$. 
A rapid approach to a slow evolution of the equation of state 
in classical field
theories has also been observed with expanding space-time 
geometries~\cite{Borsanyi:2003ib}. 
The fast loss of memory for these quantities 
is based on the phenomenon of ``dephasing''~\cite{Cooper:1996ii}, which is 
independent of the interaction details.
In contrast, ``mode temperatures'' (to be defined below) for individual
momentum modes loose only part of the initial condition details
on a somewhat longer time scale $t_{\rm damp}$ which depends on the
interaction strength~\footnote{This partial loss of memory 
can be related to approximate ``fixed point'' solutions for  
time evolution equations of equal-time correlators~\cite{Bettencourt:1997nf}.}.
Still, $t_{\rm damp}$ is much smaller than
the true thermal equilibration time $t_{\rm eq}$ which describes
the universal rate of approach to the equilibrium values for
all relevant correlation 
functions~\cite{Berges:2000ur,Cooper:2002qd,Berges:2002wr}.
An even substantially larger separation of scales is observed
in classical field theories~\cite{Aarts:2000mg,Borsanyi:2000pm,Salle:2000jb}
as compared~\cite{Aarts:2001yn} to the corresponding quantum theories. 

\begin{figure}[t]
\begin{center}
\vspace*{0.cm}
\hspace*{0.cm}
\epsfig{file=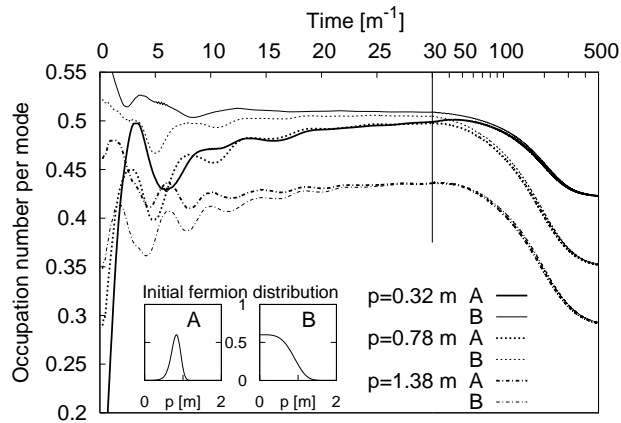,width=8.5cm}
\end{center}
\vspace*{-0.7cm}
\caption{Fermion occupation number $n^{(f)}(t;p)$
for three different momentum modes as a function of time. 
The evolution is shown for two different initial
conditions with {\em same} energy density. 
The long-time behavior is shown on a 
logarithmic scale for $t \ge 30\, m^{-1}$.}
\label{fig:join_fn}
\end{figure}
In this letter we consider the nonequilibrium evolution of
quantum fields for a low-energy  
quark-meson model, which takes into account 
two quark flavors with a Yukawa coupling $\sim h$ 
to a scalar $\sigma$-field and a triplet of pseudoscalar pions, $\vec{\pi}$. 
The theory corresponds to the well-known ``linear $\sigma$-model'',
which incorporates the chiral symmetries of massless two-flavor QCD.  
The classical action reads
\bea
S &=& \int {\rm d}^4 x \Big\{\bar{\psi} i \partial\!\!\!\slash \psi 
+\frac{1}{2}\left[\partial_\mu \sigma \partial^\mu \sigma
+ \partial_\mu \vec{\pi}  \partial^\mu \vec{\pi} \right] \nonumber\\
&& \qquad\quad\!\!\!  
+\, h \bar{\psi} \left[\sigma + i\gamma_5 \vec{\tau} \vec{\pi} \right] \psi
- V(\sigma^2 + \pi^2) \Big\} \, .
\label{chiralfermact}
\eea
We consider a quartic scalar self-interaction 
$V(\sigma^2 + \pi^2) = m_0^2 \left(\sigma^2 + \pi^2\right)/2 
+ \lambda \left(\sigma^2+ \pi^2\right)^2/(4! N_f^2)$ with $N_f=2$.
The employed couplings are taken to be of order one, and if
not stated otherwise $h=\lambda=1$.
We emphasize that the main results of this letter about
prethermalization are independent
of the detailed values of the couplings.
Here we use the two-particle irreducible (2PI) effective action to 
two-loop order~\footnote{With the rescaling $h \to h/N_f$ this corresponds
to a nonperturbative expansion of the 2PI effective action to 
next-to-leading order in $N_f$~\cite{Berges:2000ur,Berges:2002wr}.}.
In Ref.~\cite{Berges:2002wr} it has been shown that this approximation can be
used to study the far-from-equilibrium dynamics as well as the 
late-time approach to quantum thermal equilibrium. The
dynamics is solved numerically without further approximations
(cf.~Ref.~\cite{Berges:2002wr} for calculational details). All quantities
will be given in units of the scalar thermal
mass $m$ \footnote{The thermal mass $m$ is evaluated in equilibrium.
It is found to prethermalize very rapidly. The employed momentum cutoff
is $\Lambda/m = 2.86$.}.  

\begin{figure}[t]
\begin{center}
\vspace*{0.cm}
\hspace*{0.cm}
\epsfig{file=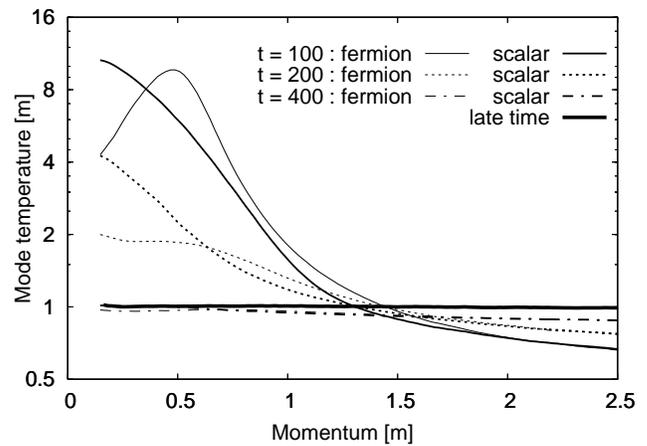,width=8.5cm}
\end{center}
\vspace*{-0.7cm}
\caption{Fermion and scalar mode temperatures $T^{(f,s)}_p(t)$ as 
a function of momentum $p$ for various times.}
\label{fig:Tdist}
\end{figure}
{\em Thermalization:} Nonequilibrium dynamics requires the
specification of an initial state. A crucial question of thermalization is
how quickly the nonequilibrium system effectively
looses the details about the initial conditions, and 
what are the characteristic stages of a partial loss of information.
Thermal equilibrium keeps no memory about 
the time history except for the values of a few conserved charges.
In Fig.~\ref{fig:join_fn} we show the effective occupation
number density of fermion momentum modes, $n^{(f)}(t;p)$, 
as a function of time for three different 
momenta~\footnote{This quantity is directly related to the expectation value
of the vector component of the field commutator 
$\langle [\psi,\bar{\psi}] \rangle$ in Wigner
coordinates and fulfills $0 \le n^{(\rm f)}(t;p) \le 1$~\cite{Berges:2002wr}.}.
The plot shows two runs denoted as (A) and (B) 
with different initial conditions but same energy density.
Run (A) exhibits a high initial particle
number density in a narrow momentum range around
$\pm p$. This situation is reminiscent of  
two colliding wave packets with opposite 
and equal momentum. We emphasize, however, that we 
are considering a spatially homogeneous and isotropic
ensemble with a vanishing net charge density.   
For run (B) an initial particle number density is employed which
is closer to a thermal distribution. 

One observes that for a given momentum the mode numbers
of run (A) and (B) approach each other at early times. 
The characteristic time scale for
this approach is well described by the damping
time $t_{\rm damp}(p)$~\footnote{The rate $1/t_{\rm damp}(p)$ is 
determined by the spectral component of the 
self-energy~\cite{Berges:2002wr}.}. 
Irrespective of the initial distributions (A) or (B), we find 
(for $p/m\simeq 1$)
$t_{\rm damp}^{(f)} \simeq 25\, m^{-1}$ for fermions
and $t_{\rm damp}^{(s)} \simeq 28\, m^{-1}$ for 
scalars. In contrast to the initial rapid changes, 
one observes a rather slow or ``quasistationary'' subsequent
evolution. The equilibration time
$t_{\rm eq} \simeq 95\, m^{-1}$ 
is substantially larger than
$t_{\rm damp}$ and is approximately the same for fermions and 
scalars. Thermal equilibration is a collective phenomenon which is,
in particular, rather independent of the momentum.
In summary, mode quantities such as effective
particle number distribution functions show a characteristic 
two-stage loss of initial conditions: after the 
damping time scale much of the 
details about the initial conditions are effectively lost. 
However, the system is still far from equilibrium and  
thermalization happens on a much larger time scale.

\begin{figure}[t]
\begin{center}
\vspace*{0.cm}
\hspace*{0.cm}
\epsfig{file=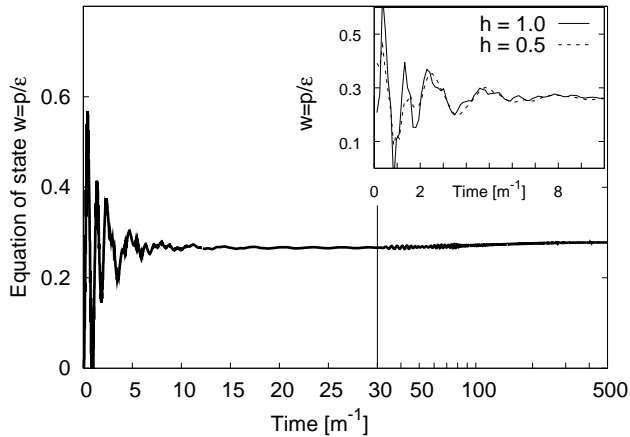,width=8.5cm}
\end{center}
\vspace*{-0.7cm}
\caption{The ratio of pressure over energy density $w$ as a function of
time. The inset shows the early stages 
for two different couplings and demonstrates
that the prethermalization
time is independent of the interaction details.}
\label{fig:wevol}
\end{figure}
We define mode temperatures $T^{(f,s)}_p(t)$
by equating the mode particle numbers $n^{(f,s)}_p(t)$ with a 
time and momentum dependent Bose-Einstein or Fermi-Dirac distribution, 
respectively~\cite{Berges:2002wr}:
\beq
n_p(t) \stackrel{!}{=} 
\left[ \exp\left(\omega_p(t)/T_p(t) \right) 
\pm 1 \right]^{-1} \, .
\label{eq:occup}
\eeq
This definition is a quantum mechanical
version of its classical counterpart as defined by the
squared ``generalized velocities''~\cite{Bonini:1999dh}.
In thermal equilibrium with $\omega_p \simeq \sqrt{p^2 + M^2}$
and $T_p = T_{\rm eq}$
Eq.~(\ref{eq:occup}) yields the familiar occupation numbers ($\mu=0$).
Here the mode frequency $\omega^{(f,s)}_p(t)$ is determined by the peak of the
spectral function for given time and momentum, as detailed in
Ref.~\cite{Berges:2002wr}. In Fig.~\ref{fig:Tdist} we show the fermion and 
scalar mode temperature as a function of momentum for various 
times $t \gg t_{\rm damp}$. 
One observes that at late times, when thermal equilibrium is approached,
all fermion and scalar mode temperatures become constant 
and agree: $T_p^{(f)}(t) = T_p^{(s)}(t) = T_{\rm eq}$. In contrast,
there are sizeable deviations from the thermal result
even for times considerably larger than the characteristic damping 
time.  

\begin{figure}[t]
\begin{center}
\vspace*{0.cm}
\hspace*{0.cm}
\epsfig{file=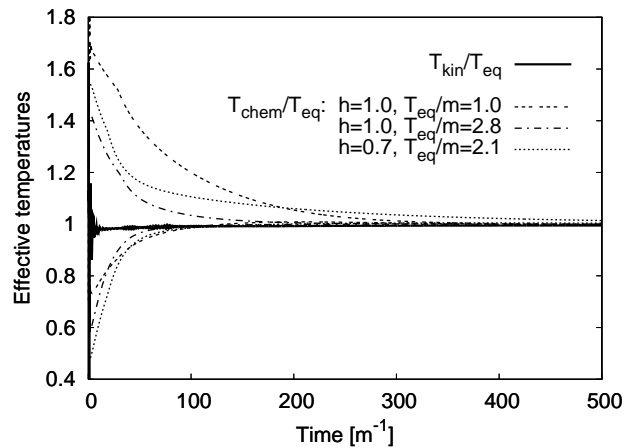,width=8.5cm}
\end{center}
\vspace*{-0.7cm}
\caption{Chemical temperatures for scalars (upper curves) and
fermions (lower curves) for different 
values of the coupling $h$ and $T_{\rm eq}$.
We also show the kinetic temperature $T_{\rm kin}(t)$ (solid line), 
which prethermalizes on a very short time scale as compared to
chemical equilibration.}
\label{fig:Tkinevolrat}
\end{figure}
{\em Kinetic prethermalization:}
In contrast to the rather long thermalization time,
prethermalization sets in extremely rapidly. 
In Fig.~\ref{fig:wevol} 
we show the ratio of pressure over energy density,
$w = p/\epsilon$, as a function of time. One observes that
an almost time-independent equation of state builds up 
very early, even though the system is still far from equilibrium!
The prethermalization time $t_{\rm pt}$ is here 
of the order of the characteristic
inverse mass scale $m^{-1}$. This is a typical consequence of
the loss of phase information by summing over oscillating functions 
with a sufficiently dense frequency spectrum. In order to see that this 
phenomenon is not related to scattering or 
to the strength of the interaction, we compare
with a smaller coupling in the inset and observe good agreement
of both curves. The dephasing phenomenon is 
unrelated to the scattering-driven process of thermalization.

Given an equation of state, the question arises 
whether there exists a suitable definition of a global 
kinetic temperature $T_{\rm kin}$. In contrast to a mode quantity such 
as $T_p(t)$, a temperature
measure which averages over all momentum modes may prethermalize.
Building on the classical association of temperature with
the mean kinetic energy per degree of freedom, we use here
a definition based on the total kinetic 
energy $E_{\rm kin}(t)$:
\beq
T_{\rm kin}(t) = E_{\rm kin}(t)/c_{\rm eq}\,  . 
\eeq
Here the extensive dimensionless proportionality constant  
$c_{\rm eq} = E_{\rm kin,eq}/T_{\rm eq}$ is given 
solely in terms of equilibrium
quantities~\footnote{For a relativistic plasma
one has $E_{\rm kin}/N = \epsilon/n = \alpha T$.
As alternatives, one may consider the weighted average
$\bar{T}(t) = \sum n(t;p) T(t;p)/ \sum n(t;p)$ where the sum is over 
all modes, or a definition analogous to Eq.~(\ref{eq:chem}).}. 
Since total energy is conserved, the time scale 
when ``equipartition'' is reached (i.e.~$E_{\rm kin}/E$ is 
approximately constant) also corresponds to a time-independent 
kinetic temperature. The latter equals the equilibrium temperature
$T_{\rm eq}$ if $E_{\rm kin}/E$ has reached the thermal value.

The solid line of Fig.~\ref{fig:Tkinevolrat} shows $T_{\rm kin}(t)$
normalized to the equilibrium temperature (for $T_{\rm eq}/m = 1$). 
One observes that an almost time-independent kinetic temperature is
established after the short-time scale $t_{\rm pt} \sim m^{-1}$. 
The time evolution of bulk quantities such
as the ratio of pressure over energy density $w$, or the
kinetic temperature $T_{\rm kin}$, are dominated by
a single short-time scale. These quantities approximately
converge to the thermal equilibrium values already at early
times and can be used for an efficient ``quasi-thermal'' 
description in a far-from-equilibrium situation!

{\em Chemical equilibration:}
In thermal equilibrium the relative particle numbers of different species 
are fixed in terms of temperature and particle masses. A system
has chemically equilibrated if these ratios are reached, as observed for 
the hadron yields in heavy ion collisions~\cite{Braun-Munzinger:2003zd}.
Obviously, the chemical equilibration time $t_{\rm ch}$ will
depend on details of the particle number changing interactions in a 
given model and $t_{\rm ch} \le t_{\rm eq}$. In our
model we can study the ratio between the numbers of fermions
and scalars. For this purpose we introduce the 
chemical temperatures $T_{\rm ch}^{(f,s)}(t)$ 
by equating the integrated number density
of each species, $n^{(f,s)}(t) = g^{(f,s)}
\int {\rm d}^3 p/(2 \pi)^3\, n^{(f,s)}_p(t)$, 
with the integrated Bose-Einstein/Fermi-Dirac form of distributions:
\beq
n(t) \stackrel{!}{=} 
\frac{g}{2 \pi^2} \int_{0}^{\infty}\! {\rm d} p p^2
 \left[ \exp\left(\omega_p(t)/T_{\rm ch}(t) \right) 
\pm 1 \right]^{-1} \, .
\label{eq:chem}
\eeq      
Here $g^{(f)}=8$ counts the number of fermions and 
$g^{(s)}=4$ for the scalars. 

The time evolution of the ratios 
$T_{\rm ch}^{(s,f)}(t)/T_{\rm eq}$ is shown
in Fig.~\ref{fig:Tkinevolrat} 
for different values of the coupling constant $h$ and  
the equilibrium temperature $T_{\rm eq}$. 
One observes that chemical equilibration with 
$T_{\rm ch}^{(s)}(t) = T_{\rm ch}^{(f)}(t)$ does not
happen on the prethermalization time scale, in contrast
to the behavior of $T_{\rm kin}(t)$.
Being bulk quantities, the scalar and fermion 
chemical temperatures can approach each other rather quickly at first. 
Subsequently, a slow evolution towards equilibrium sets in.   
For the late-time chemical equilibration we find for
our model $t_{\rm ch} \simeq t_{\rm eq}$. However, the
deviation from the thermal result can become relatively small
already for times $t \ll t_{\rm eq}$. 

Let us finally consider our findings in view of collisions of heavy
nuclei and try to estimate the prethermalization time. Actually,
$t_{\rm pt}$ is rather independent of the details of the model
like particle content, values of couplings etc. It mainly reflects
a characteristic frequency of the initial oscillations. 
If the ``temperature'' (i.e.~average kinetic energy per mode)
sets the relevant scale one expects
$T\, t_{\rm pt} = {\rm const}$.
(For low $T$ the scale will be replaced by the mass.)
For our model we indeed find 
$T\, t_{\rm pt} \simeq 2 - 2.5$~\footnote{We
define $t_{\rm pt}$ by $|w(t_{\rm pt}) - w_{\rm eq}|/w_{\rm eq}
< 0.2$ for $t > t_{\rm pt}$.}. We expect
such a relation with a similar constant to hold for the quark-gluon
state very soon after the collision~\footnote{To establish
this would require the application of
similar techniques to QCD~\cite{Berges:2004pu}.}.
For $T \gtrsim 400 - 500\,$MeV we
obtain a very short prethermalization
time $t_{\rm pt}$ of somewhat less than $1\,$fm. This is consistent 
with very early hydrodynamic behavior~\footnote{Future work should
investigate isotropization of pressure.}.
In QCD the equilibrium 
equation of state shows no strong temperature dependence
above the critical temperature $T_{\rm c}$~\cite{Karsch:2003jg}, 
and can therefore
adapt easily as the temperature decreases. After the
transition $w$ will only re-adjust somewhat to the
equilibrium value relevant for a hot hadron gas, typically
on a time scale of a few fm (for $T \simeq 175\,$MeV). 
The chemical equilibration time $t_{\rm ch}$ depends on the production rate for
multi-strange hadrons~\cite{Braun-Munzinger:2003zz}. 
From $t_{\rm pt} \ll t_{\rm ch}$
and Fig.~\ref{fig:Tkinevolrat} we conclude that once the chemical
temperatures for the various different species are equal
the relevant chemical temperature $T_{\rm ch}$
coincides with $T_{\rm kin}$ and defines a universal temperature.
Comparison with the critical temperature in equilibrium 
is therefore meaningful -- an approximate equality 
$T_{\rm ch} \simeq T_{\rm c}$ has been 
advocated~\cite{Braun-Munzinger:2003zz} --
such that chemical freeze out can, in principle, be used
to measure $T_{\rm c}$.

We thank J.~Serreau for helpful discussions and collaboration on
related work.

\end{document}